\documentclass[runningheads]{cl2emult}
\usepackage{amsmath}

\newcommand{\myel}{{\mathcal L}}

\newcommand{\nei}{{\mathcal N}}
\newcommand{\well}{{\mathcal W}}

\newcommand{\prob}{{{\mathcal M}_1}}
\newcommand{\ee}{{\bf E}}
\newcommand{\goesto}{\rightarrow}
\newcommand{\amin}{\arg\min}

\newcommand{\gee}{{\mathcal G}}
\newcommand{\dee}{{\mathcal D}}
\newcommand{\eps}{\epsilon}

\newcommand{\kinf}{\lim_{k \goesto \infty}}

\newcommand{\cspace}{\mbox{, }}

\newcommand{\other}{\mbox{\rm {otherwise}}}

\newcommand{\te}{{\tau(\eps)}}
\newcommand{\X}{{\mathcal X}}
\newcommand{\finv}{{f^{-1}}}
\newcommand{\dinv}{{d^{-1}}}

\newcommand{\minf}{\left( \left. \left[M, \infty \right. \right) \right)}

\newcommand{\avmomgen}{{\ee^\ast \left[ e^{ \xi \te} \right]}}

\newcommand{\ejx}{{e^{j\xi}}}

\newcommand{\pjejx}{{p^j \ejx}}

\newcommand{\peemu}{{{\rm P}^\ast_p}}

\newcommand{\qj}{{q(j)}}
\newcommand{\sumj}{\sum_{j=0}}

\input{epsf}

\begin{document}

\title*{Evolution at the Edge of Chaos: A Paradigm for the 
Maturation of the Humoral Immune Response}
\titlerunning{Evolution at the Edge of Chaos}

\author{Patricia K. Theodosopoulos \and Theodore V. Theodosopoulos}
\institute{}


\maketitle

\begin{abstract}
We study the maturation of the antibody population following 
primary antigen presentation as a global optimization problem.  
Emphasis is placed on the trade-off between the safety of 
mutations that lead to {\it local} improvements to the antibody's 
affinity and the necessity of eventual mutations that result in 
{\it global} reconfigurations in the antibody's shape. The model 
described herein gives evidence of the underlying optimization 
process from which the rapidity and consistency of the biologic 
response could be derived.  
\end{abstract}

\section{Introduction}

The study of the mechanisms underlying the physiology of the 
immune system has been a very promising area for applications of 
mathematical models.  The spectacular success of the healthy 
immune system to recognize the combinatorial plethora of antigenic 
agents while being endowed with a substantially smaller repertoire 
of immunoglobulin (Ig) molecules is a problem that shares much in 
common with mathematical problems of combinatorial complexity and 
optimization.

This article focuses on the maturation of the humoral immune 
response as a conveniently fast-paced example of selection-based 
evolution.  Through this example, we attempt to investigate the 
interplay between lessons learned from mathematical models of 
global optimization and the need for theoretical models in the 
biological sciences. 

The paper begins by providing a brief overview of the B cell 
immune response.  The following sections provide the biological 
context for the presentation of our model and results, and a 
summary of the techniques involved in our analysis.  The 
presentation of our model and results proceeds in five steps of 
increasing complexity which assume a familiarity with the 
corresponding sections in the biological discussion of the model.

\section{The Naive B Cell Repertoire}

The ability of the specific immune response to recognize and 
respond to myriad foreign antigen challenges rests in the 
generation of diversity at different stages in the development of 
the immune cells.  With regards to the B cell, diversity is first 
explored at the time of antibody naive maturation with 
somatic recombination from a pool of genes that construct the 
variable and constant domains of the heavy and light chains of the 
antibody molecule, along with their inherent combinatorial 
freedom.  At this stage, there is additional diversity introduced 
via imprecise rearrangements and junctional nucleotide insertions. 
It is now believed that conformational isomerism of Ig molecules 
may also add additional diversity to the primary repertoire 
\cite{foote1}. Conservatively, these processes have been 
calculated to generate a repertoire on the order of 
$10^9$ \cite{elgert}.

\section{B Cell Response}

The introduction of foreign antigen into the host results in a 
complex response that occurs rapidly and effectively.  The initial 
phase following antigen introduction  involves elimination via 
innate immunity.  The  mediators of this response are nonspecific, 
including the phagocytic cells, complement, and NK cells.  
Subsequently, the mediators of specificity in the immune response, 
represented by the T cells and B cells, are activated following 
interactions with macrophages, and other soluble factors.  The 
evolution of specificity into the immune repertoire greatly 
enhanced the organism's ability to respond to foreign invaders, 
and even more importantly, to develop memory of this invader that 
is protective upon reintroduction of the pathogen.

Antigen localization following exposure occurs within two 
compartments in the lymph node. The first is the primary 
follicles, which are comprised of antigen--antibody complexes and 
follicular dendritic cells that present antigen to a circulating 
population of B cells. The second compartment is the 
paracortex.  Within minutes of introduction, the antigen is 
taken up by the phagocytic cells present in the paracortex. These cells will 
process the antigen for presentation via MHC to specific CD4+ T 
cells. The interaction of the T cells, antigen-presenting cells 
(APCs), and B cells will give rise to a population of low affinity 
B cells that can generate large amounts of antibody for a fixed 
period of time, called the plasmacytes of the primary response, and 
another population that will give rise to the secondary follicles 
and germinal centers (GC) where hypermutation will take place 
\cite{elgert,berek}.  Usually only 3--5 of these antigen 
stimulated B cells enter a given follicle and there they undergo 
exponential growth that fills the follicular network in 3--4 days 
\cite{berek}. These cells, called centroblasts, are 
believed to lose surface Ig expression and undergo hypermutation. 
It is believed that these hypermutating cells undergo several 
rounds of mutation followed by selection within the 
microenvironment of the germinal center \cite{kepler,celada}.  
The selection process is believed to encompass 
both a positive selection for higher affinity and a negative 
selection barrier to remove clones that have developed self-recognizing 
phenotype or other detrimental mutations \cite{hande}.  
Those cells that pass the selection barriers enter the circulation 
as high affinity plasma cells or memory cells. 

\section{The Process of Hypermutation}

The primary repertoire appears to be sufficient for the organism 
to recognize with a certain threshold affinity, and in some cases 
even high affinity, the antigenic challenges presented by the 
environment. The additional mechanism of somatic hypermutation, 
that in humans appears to be primarily antigen driven, improves 
the affinity of the antibody for the antigen by two orders of 
magnitude or more, with some expense of energy and cells, and some 
risk (i.e., autoimmunity and malignancy) to the organism.  It has 
been hypothesized that this process is simply an evolutionary 
relic \cite{blanden} that was initially needed to generate 
primary repertoire locally.  However, it seems unlikely that this 
process which produces immunologic memory and high affinity 
effector antibody molecules at some expense and risk to the 
organism is redundant.  It provides at the very least the security 
that in a world of countless and evolving pathogens, the organism 
can protect itself, but one might also suspect that the true 
utility of this process may even extend beyond antigen response 
and into a larger control mechanism for the organism, whose role 
may be appreciated more during early development, \cite{schroeder} 
or during the shaping and maintenance of memory.

Although the actual mechanism of hypermutation is still not 
understood, it appears to have some link to the transcription 
process,  and several models have been suggested along these lines 
\cite{fukita,klix,diaz,blanden}.  The 
mutations introduced are primarily point mutations, although 
deletions and insertions do occur, and more frequently than 
previously suspected \cite{klein}. Mutations seem to occur 
preferentially in the region bounded by the transcriptional 
promoter at the 5$'$ end, and the C gene at the 3$'$ end. The pattern 
displayed is that of a rapid peak in mutation frequency, followed 
by a slow decline out to about 1.5--2kb downstream 
\cite{elgert,lebecque}.  The regions of both light and heavy chain V 
genes that are selected have an average of 3--13 mutations, but can 
have upwards of 20 \cite{green}. The primary targets for 
hypermutation are the CDRs, or complementary determining regions, 
of which there are three in both heavy and light chain, separated 
from each other by intervening framework sequences(FRW).  The 
CDRs are only a few residues in length but their position in the 
protein molecule and configuration in three-dimensional space make 
them crucial in the evolution of diverse antigen combining sites 
\cite{elgert}.  

The substrates within each CDR that are frequently seen mutated 
are defined as ``hotspots''.  They are described by preferences 
for purines, rather than pyrimidines, as well as for particular 
codons, or codon motifs within the sequence.  The fact that 
mutation in a {\it hotspot} can create or delete other hotspots 
indicates a higher order structure to the mutation process than 
that which is currently observable \cite{brown}. In vitro random 
mutagenesis studies show loss of around 50\% of clones that 
accumulate more than one mutation \cite{wiens}.  This is due to 
the effects of both diminished antigen binding, as well as loss of 
expression of a functional Ig molecule.  These cells are believed to 
undergo apoptosis, perhaps mediated via T cells 
\cite{liu,andersson}.   Despite the evidence 
suggesting high loss and apoptosis in germinal centers 
\cite{liu}, true numbers of in vivo loss are not well 
documented.  If mutation results in the production of a functional 
Ig molecule, then it is believed to be tested for affinity 
against the available antigen trapped in the follicular dendritic 
network of the particular germinal center.  The role of 
competition for limited antigen, although figuring prominently in 
prior models \cite{kepler}, is still being elucidated 
\cite{manser}. Following this process of selection, which appears 
to have both a positive and a negative barrier as described above, 
the high affinity antibody-producing B cell may leave the germinal 
center and enter the circulation as a plasma cell or a memory cell 
\cite{hande}.

\section{The Model}

The model presented here attempts to delineate a selection-based 
model of the evolution of the affinity-matured antibody.  The 
contribution of the microenvironment in the germinal center as 
well as the intrinsic properties of primary repertoire antigen--
antibody interactions versus affinity-matured interactions are 
considered.  In addition, the desire to understand such 
observations as repertoire shift of variable region genes in the 
memory compartment, as well as to suggest an underlying mechanism 
of somatic hypermutation which is an unique adaptive evolutionary 
process in mature organisms are considered.  The following 
sections give a more detailed biological context within which the 
methods and results can be interpreted.

\subsection{Local Steps versus Global Jumps}

This component attempts to model the biological ``trade-off'' that 
occurs during the mutation process and allows the rapid generation 
of high affinity antibodies. One might understand this trade-off 
in terms of the mutations that produce only local changes in the 
conformation and are therefore more likely, although not 
exclusively, to produce incremental changes in the affinity, 
versus those mutations that produce more global changes in 
conformation and therefore might be expected to produce rather 
large {\it jumps} in affinity.  

In terms of affinity, the prior treatment of affinity changes and 
mutational studies lead to the idea that through the mutation 
process, the selection is for those clones that undergo a stepwise 
increase in affinity \cite{nossal,brown} -- an additive 
effect of changes that create new H bonds or new weak 
electrostatic or hydrophobic interactions between the residues and 
associated solvent molecules \cite{covell,andersson,braden}.  
However, it is observed that all codon changes 
cannot be translated into stepwise energetic changes 
\cite{covell}. In the literature, this  affinity increase is often 
correlated with a lower $K_{off}$ more so than a higher 
$K_{on}$ for affinity measurements, although which one is more 
important for overall affinity increase of Igs is still unclear 
\cite{wedemayer,andersson}. With regards to the changes 
in conformation, the nature of the affinity change secondary to 
the stepwise energetic changes in the selected antibodies has also 
led to the idea that the conformation is handled likewise. This 
progression to a {\it lock and key} conformation occurs at the 
expense of entropy in exchange for a decrease in free energy and a 
commensurate increase in affinity \cite{wedemayer}. Since we 
cannot reliably observe the process, we cannot presume that this 
stepwise search is what is always functioning in the germinal 
center.  We predict that the rapid elaboration of high affinity 
antibodies through the germinal center reaction may necessitate that
the system occasionally make a large {\it jump} in order to 
better sample the affinity landscape. 

The positions of positively selected mutations show that 
replacement mutations occur preferentially in the CDRs versus the 
intervening framework regions (FRWs). The FRW was often 
described as being very sensitive to replacement mutations, but it 
appears now that they too can tolerate a certain number of 
replacement mutations, and that the CDRs may alternately, possess 
a sensitivity to mutation through the coding structural elements 
as well \cite{wiens}. The greatest diversity a priori is seen in 
the CDR 3, which appears to have the most contact residues with the 
antibody, while the other CDRs usually comprise the sides 
of the binding pocket \cite{schroeder}.  During hypermutation, 
it is often in CDR 1 and 2 that one observes most of the 
mutations, whereas in CDR 3, there are relatively fewer, and they 
do not usually affect the existing contact residues 
\cite{schroeder,dorner,tomlinson}. We might then 
hypothesize that the local steps will result preferentially from 
mutations in CDR 1 and 2 and that global conformation changes 
might occur from CDR 3 or even FRW mutations.  This is of course 
not absolute, as experimentally, CDR 2 regions have also been seen 
to contribute to the binding pocket, and to have long range 
interactions at certain residues that make mutations in them 
change significantly the antibody conformation \cite{wiens}).  
Furthermore, certain base pair positions that are frequently 
mutated in the CDRs may create conservative local changes, while 
mutations outside of these positions may be more frequently 
associated with global changes \cite{tomlinson,lebecque}.  
The less frequently mutated codons are more 
common within the CDR 3, and this CDR experiences fewer mutations 
than the other CDRs or FRW regions, thus supporting the above 
generalization. 

Growing evidence suggests that the primary repertoire is composed 
of multivalent, and highly flexible Igs that {\it conform} to the 
antigen, but through hypermutation they generate a rigid {\it lock 
and key} fit. Studies comparing germline diversity with 
hypermutated V genes showed that the amino acid differences 
introduced by mutation were fewer than the underlying diversity of 
the primary repertoire, and further suggested that through 
mutation, the more conserved residues of the CDR 1 and 2 that 
often create the periphery of the binding site are favored for 
mutation, whereas the more diverse residues are generally not 
mutated \cite{dorner}.  Although exceptions were cited for both 
of these generalizations, this supports a notion that the mutation 
mechanism has evolved to focus mutations primarily in those 
residues which, in the three-dimensional geometry of the CDR loops, 
will create a tighter fit for the antigen, and decrease 
flexibility of CDR 1 and CDR 2 at the periphery of the binding 
pocket \cite{tomlinson}. The diversity of the primary repertoire 
can rarely achieve this energetic feat. 

An additional mechanism for global jumps may be appreciated from 
the relative frequency of deletions and insertions.  Recent work 
has identified deletions and insertions in single cell analysis of 
GC derived B cells at a frequency much higher than previously 
suspected, in the range of 4--16 percent of in-frame rearrangements 
\cite{klein}.  These types of alterations would be suspected to 
contribute greatly to the occurrence of global changes in 
conformation and affinity.

\subsection{Consistency of Response}

Another component of the model will be the robustness of this 
evolutionary optimization to the diverse population of antigens 
that the organism faces. How do the mutation and selection 
processes ensure consistency in their response?  The resulting 
affinities as well as the timing of response exhibit remarkable 
consistency both within the response to a given antigen and across 
responses to diverse antigens.  The naive repertoire usually 
produces antibodies with affinities on the order of $10^5 M^{-1}$, 
and the somatic hypermutation process produces antibodies with a 
range of affinities from $10^6$ to $10^8 M^{-1}$ within a period 
of days from a finite number of clones. 

Additional consistency arises within a response to a given 
antigen, as the high affinity clones share many favorable 
mutations \cite{manser}.  Although other amino acid 
substitutions in that same position also confer high affinity, 
they are not selected for or observed in the mature response 
\cite{manser}.  This is considered as evidence of a negative 
selection barrier in the process that might be protecting against 
harmful mutations \cite{manser,hande}, although it may 
be an intrinsic feature of the mutating sequences. There is also 
evidence of consistency in specific base pair positions selected 
for mutation across different responses in both productive and 
non-productive rearrangements \cite{dorner}.  The codons in the 
CDRs seem, by their nature, to be predisposed in favor of 
replacement mutations during hypermutation.  How does the system 
work within the constraints of the available number of clones, the 
timing observed for response, and even the physical or energetic 
limitations of the mechanism to ensure a consistent response? 
\cite{manser2}. The potential benefits of the germinal center 
may be discussed in this respect. 

The compartmentalization of the germinal center provides ease of 
interaction of the necessary components, ability to segregate 
beneficial from detrimental mutations in a controlled fashion, 
perhaps decreased energy expenditure for the organism, and improved 
diversity of antibodies, since each germinal center appears to 
function autonomously \cite{manser}.  If we treat our model 
according to this compartmentalized representation, the parameters 
are defined independently within each germinal center. The initial 
number of clones entering a given GC is observed to be 3--5 antigen stimulated 
B cells that have received helper T cell signals \cite{elgert}.  
Previous models have suggested that these cells enter a 
division phase that fills the GC over a period of 3--4 days 
\cite{maclennan,kepler}, 
stop expressing surface Ig, and begin hypermutation. This 
would be followed by selection against the antigen trapped within 
the individual GC follicular dendritic network. Current evidence 
favors several rounds of mutation and selection \cite{kepler}, 
and more than one mutation per cycle of hypermutation 
\cite{manser2,goyenechea}.  In between these rounds of 
selection and mutation, one might presume that the positively 
selected clones would have to be given an advantage, i.e., by 
generating more offspring, but how this is translated into a number 
of cell divisions is not clear, nor is how the final decision is 
made to allow the cell to exit the GC when and if it attains a 
high enough affinity.  Maybe it is a time-related phenomenon, or 
perhaps an affinity threshold controlled process.   Even with the 
benefit of the fastest observed generation time in the GC, which is 
6 hours \cite{manser2}, the 3--4 days of generation time gives a 
population of $10^4$-$10^5$.  From this starting point, we want 
to understand the probability of generating a high affinity clone, 
and how many mutations it will take to get us there.

\subsection{The Affinity Threshold}

A third component will be to treat the dependence of response time 
on the affinity threshold level.  Following the initial antigen 
presentation, at around day 5, there appear in circulation low 
affinity plasma cells.  The germinal center begins to form a few 
days after antigen exposure, and the first mutated V genes are 
detected around day 5--7 \cite{manser2,song}. Samples of 
V genes throughout this primary GC reaction show increasing 
numbers of mutations over time \cite{manser2}. Subsequent 
immunization produces a fast response of high affinity antibodies, 
usually within 1--3 days \cite{elgert}.  The V genes of the high 
affinity antibodies of secondary and tertiary immune responses are 
also seen to have more accumulated mutations, but the incremental 
increase in affinity following the primary response is low 
\cite{elgert,foote2}. This implies a point of 
diminishing returns for this process. 

Interestingly, the high affinity antibody molecules of the 
secondary response frequently use different V genes than the 
primary response population.  This is referred to as {\it  repertoire 
shift} and there does not appear to be enough time for these cells 
to be created de novo from the newly forming germinal centers.  
Therefore, these cells must have evolved into memory either late 
in the primary germinal center response or during the interval 
between. This of course presumes adequate time between 
innoculations, as too short or long an interval between exposures will produce a 
diminished response. In humans, hypermutation seems to be largely 
confined to the germinal center.  The reaction in the GC lasts 
about 2--3 weeks, although antigen has been shown to remain on 
follicular dendritic cells for years following the primary immune 
response \cite{elgert} and Ig-expressing B cell blasts can be 
evident for months following initial exposure \cite{maclennan}. 
It has been suggested that this remaining antigen is the force 
behind the shaping of the memory immune response between 
exposures, and that it also accounts for the repertoire shift.  In 
addition, the memory compartment for a given epitope is often 
oligoclonal, whereas the high affinity late primary stage 
clonotypes can be numerous \cite{manser}. It is possible that 
following the primary exposure, there are two populations of cells 
generated through the germinal center-- a fast response effector 
population and a slow response memory population. It is 
interesting to note that there may be an energetic advantage to 
some of these repertoire shifted memory cells--that they do not 
necessarily have higher affinity, but perhaps they cross an energy 
barrier more easily \cite{foote2}.  This would imply kinetic 
selection as opposed to affinity selection during this phase of 
the immune response. Attempting to understand these other 
selection parameters in shaping the memory response would be an 
interesting prospect for future models.

\subsection{Evolution of the Mutation Dynamics}

The fourth component deals with the evolution of the above 
outlined trade-off and how it is optimized in the organism. The observed 
mutation distribution is not random, because if it were, the 
degeneracy rate predicted would be too high. Instead it has 
evolved to target certain sites preferentially 
\cite{manser,dorner}. Intrinsic hotspots are usually 
characterized as 3--4 base pair sequences such as AGC 
and TAC and their inverted 
repeats and RGYW motifs (R=purine, Y=pyrimidine and W=A,T) 
\cite{yelamos,dorner2}.  These motifs are shown to be 
hotspots independent of antigenic selection, and concentrated in 
the CDRs, often in an overlapping fashion \cite{dorner2,elgert}. 
Other characteristics of this mutator mechanism 
include a preference for point mutations over 
deletions/insertions, a bias against mutations in thymidine and 
in favor of mutations in purines instead of pyrimidines.  
Mutations are favored in non-degenerate sites, and the 
replacement/silent mutation (R/S) ratios are higher in the CDRs 
in both selected and non-selected V genes, also suggesting 
intrinsic targeting of these areas \cite{dorner2}. Most 
significantly, in experiments with a light chain transgene, silent 
mutation in one part of the gene resulted in loss of a hotspot 
motif and in the appearance  and loss of hotspots in other areas 
\cite{goyenechea}. This argues for a higher order template as 
well as an evolving dynamic with loss and acquisition of 
mutations. This higher order structure may be conferred by DNA 
folding, or perhaps DNA--protein interactions \cite{goyenechea}.  
Nevertheless, this dynamic might be expected to exhibit some 
convergence in order to maintain the consistency observed between 
individual responses.


\section{The Evolutionary Landscape}

As described in the previous sections, we treat the affinity
maturation of the primary humoral immune response as a problem of
global optimization.  This paradigm should be contrasted with the
``population dynamics'' approach.  The latter class of models entails
the tallying of individual immune cell types and the investigation of
the transition dynamics between their allowable states.  Such models
stress the emergence of affinity optimization as a result of these
cell population dynamics.  In this vein, it is generally the evolution
of the average affinity in the population that is the dominant
variable.

The paradigm we employ here begins instead by considering the problem
of affinity optimization.  We study the sources of complexity in this
problem and infer general principles that this optimization must
adhere to, avoiding whenever possible making ad hoc assumptions about
the particular mechanics at play.  In this regard, our model
concentrates on the hypermutation process inside the germinal centers.

Moreover, the dominant variable in our model is the maximum affinity
level that has been achieved at any stage of the hypermutation
process.  Thus, the size of the B cell population is exogenous to our
model.  We assume that the hypermutation process is initiated via a
mechanism outside the scope of our model.  Our treatment of the
hypermutation process terminates upon the development of a clone with
sufficiently high affinity.  Finally, as a result of our focus on the
affinity improvement steps, we measure time in a discrete fashion by
the inter-mutation periods.

Specifically, we begin with the space of all DNA 
sequences encoding the variable regions of the Ig molecules and a 
function on that space that models the likelihood that the 
resulting Ig molecule becomes attached to a particular antigen.  
This {\it affinity function} is conceptualized in a series of 
mappings which portray the biochemical mechanisms involved.  

To begin with, the gene in question is transcribed into RNA and 
subsequently translated into the primary Ig sequence.  This step 
describes the mapping from the genotype (a 4--letter alphabet per 
site) to the sequence of amino acids making up the Ig molecule (a 20--letter alphabet per site).  The next step is the folding of the 
resulting protein into its {\it ground state} in the presence of 
the antigen under consideration.  This step is modeled as a 
mapping from the space of amino acid sequences to the three-
dimensional geometry of the resulting Ig molecule\footnote{This 
concept is analogous to that of {\it shape space} in Chap. 13 
of \cite{rowe}.}.

Finally, the protein shape gives rise to the free energy of the Ig 
molecule in the presence of the antigen.  The free energy in turn 
is used to define the association/dissociation constants and the 
Gibbs measure which determines the likelihood of attachment.  The 
resulting affinity is visualized as a high-dimensional landscape, 
where the peaks represent DNA sequences that encode Ig molecules 
with high affinity to the particular antigen.

The process outlined above for modeling the affinity landscape 
depends on detailed knowledge which is often unavailable.  
Furthermore, our interest in the universality of the immune 
response has led us to a model that does not assume a detailed 
knowledge of the particular invading antigen.  We will return to 
this critical theme in the section on Performance Robustness.  For 
the time being, we acknowledge the need to reduce the series of 
mappings described above to a set of {\it affinity classes}, as in 
\cite{kepler}.  

The underlying theme of our modeling effort has been the 
delineation of the trade-off between the safety of mutations 
leading to {\it local steps} in shape space (and consequently 
incremental affinity improvements) and the eventual necessity of 
mutations that result in {\it global jumps} in shape space as 
discussed in section 5.1.  The main ingredient of our affinity 
landscape model 
reflects this trade-off by associating to each point in the space 
of DNA sequences the unique sequence to which it converges 
following a discrete gradient ascent algorithm.  

Let $\X$ denote the space of DNA sequences encoding the $V_H$ and $V_L$ 
regions of an Ig molecule.  Let $f$ be a positive, real-valued 
function on $\X$, which denotes the affinity function, as 
described above.  Finally, consider the gradient operator $\dee 
f(x) = \amin_{y \in \nei(x)} f(y)$, where $\{\nei(x) \subseteq \X 
\cspace \/ x \in \X\}$ describes the neighborhood 
structure\footnote{In this paper we concentrate on point mutations 
as the mechanism for local steps and thus the neighborhood we 
consider consists of all 1--mutant sequences.  It has been 
suggested \cite{manser2,goyenechea} that more than one 
point mutation may occur before the resulting Ig molecule is 
tested against an antigen presenting cell to determine its 
affinity.  Our model can capture such an eventuality by 
appropriately modifying the neighborhood structure to include the 
2-- or generally k--mutant sequences.} in $\X$.  With this notation, 
for each sequence $x \in \X$, successive applications of the 
gradient operator converge to the closest local optimum, i.e., 
there is a finite positive integer $d(x)$ and a sequence 
${\mathcal F}^\ast (x) \in \X$ such that for all $n \geq d(x)$,
$$\dee^n f(x) = \dee^{d(x)} f(x) = {\mathcal F}^\ast (x).$$
Using this association, we partition $\X$ into subsets that map to 
the same integer under $d(\cdot)$.  These level sets contain all 
sequences that are a fixed number of point mutations away from 
their closest local optimum.  

A further ingredient of our model for the affinity landscape is 
the relative nature of the separation between strictly local and 
global optima.  In practice, the global optimum is not necessarily 
the goal.  Instead, some sufficient level of affinity is desired.  
This affinity threshold is generally unknown a priori.  Our model 
allows us to view the landscape as a function of the desired 
affinity threshold.  As we show in section 9.4, we are able to 
study the dependence of our model's 
performance for a variety of affinity thresholds and thus 
investigate the trade-off between the desired affinity and the 
required time.

The consideration of strictly local versus global optima as a 
variable characteristic necessitates a finer partition of the 
level sets.  Specifically, for each level of affinity threshold, 
some of the local optima in $\X$ are below it and therefore are 
considered strictly local, while others are above it and are 
therefore considered global.  This leads to a finer decomposition 
of each level set into the part containing sequences a certain 
number of steps below a strictly local optimum versus sequences 
whose closest local optimum is also global because its affinity is 
above the desired threshold.

Let $\mu \in \prob (\X)$ be a probability 
distribution\footnote{Often, as is the case in this paper, the 
uniform distribution is used.  If there are prior preferences 
observed for particular alleles at certain base pairs, they can be 
modeled by altering $\mu$ accordingly.} on $\X$.  Let
$$\qj \stackrel{\Delta}{=} \mu \left( \well \left( \myel(M) 
\right)
\cap \dinv (j) \right)$$
and
$$p(j) \stackrel{\Delta}{=} \mu \left( \left[ \X \setminus \well 
\left( \myel(M) \right] \cap \dinv (j) \right) \right),$$
where $M$ is the affinity threshold, $\myel(M) 
\stackrel{\Delta}{=} \finv \minf$ and 
$$\well (A) \stackrel{\Delta}{=} \left \{ x \in \X :
\/ \kinf \left[\dee^k f \right](x) \in A \right \}$$
is the set of sequences that converge to a member of $A$ after 
sufficient iterations of the gradient operator.  Using this 
notation, let
$$a \stackrel{\Delta}{=} \max \left\{j \geq 0: \/ p(j) >0 
\right\}$$
and
$$b \stackrel{\Delta}{=} \max \left\{j \geq 0: \/ \qj > 0 
\right\}$$
denote the height of the strictly local and the global optima, 
respectively.  Notice that for notational convenience, we have 
suppressed the dependence of these measures on the affinity 
threshold $M$.  Unless stated explicitly otherwise, these measures 
of the affinity landscape will always depend on the affinity 
threshold as discussed above.

Despite the lack of detailed knowledge of the biologically 
relevant affinity landscapes, there is broad agreement that the 
size of the level sets decreases rapidly.  A popular model used 
previously in \cite{kepler} asserts that the level sets are 
decreasing in size at an exponential rate controlled by a 
parameter $\beta$ (in the notation of \cite{kepler} this parameter 
corresponds to $\Lambda^{-1}$).  In this paradigm, one has 
$\qj=Z_q^{-1} \beta^{-j}$ and $p(j)=Z_p^{-1} \beta^{-j}$ with 
$\sum_{j=0}^a p(j) = c \sum_{j=0}^b q(j)$, $Z_p$, $Z_q$ the 
appropriate normalization constants, and $c$, a parameter that 
determines the relative size of the set of points for which 
discrete gradient ascent traps them at strictly local optima 
versus those points for which this {\it greedy} algorithm is 
sufficient to take them to a global optimum.

While the biologically meaningful range of values for the 
parameters $a$, $b$, $c$, and $\beta$ is uncertain, the values we 
use are in agreement with the values for similar parameters used 
by other authors.  Specifically, we restrict our attention to $a,b 
\in [4,20]$, $c \in [10^4, 10^5]$, and $\beta \in [0.05,0.3]$.

\section{Optimization Dynamics}

We model the dynamics of evolutionary optimization on the affinity 
landscape as a Markov chain.  Specifically, the chain may take one 
of two actions at each time step:  it may search locally to find 
the gradient direction, and take one step in that direction or it 
may perform a global jump, which effectively randomizes the chain.  
The decision between the two available actions is taken based on a 
Bernoulli trial with probability $p$: when $p=0$, the chain 
performs global jumps all the time while $p=1$ prohibits any 
global jumps.  Thus, the parameter $p$ controls the degree of 
randomization in the Markov chain.  The biological distinction 
between local search versus global jumps is realized by means of 
at least two mechanisms described earlier: deletions/insertions 
and point replacements that lead to large changes in the resulting 
geometry. 

The mathematical description of the Markov chain model described 
above uses the following generator:
$$\left[ \gee \phi \right] (x) \stackrel{\Delta}{=} p \phi(\dee 
f(x)) + (1-p) {\bf E}^{\mu} [\phi] - \phi(x).$$
We are interested in estimating the extreme left tail of the 
distribution of the exit times for the resulting Markov chain.  In 
particular, let
$$\tau(M) \stackrel{\Delta}{=} \inf \left\{k \geq 0 | X_k \in 
\finv \minf \right\},$$
where $X_k$ denotes the Markov chain under consideration.  We are 
interested in estimating the likelihood that at least one out of a 
population of $n$ identical, non-interacting replicas of the 
Markov chain will reach an affinity level higher than $M$ before 
time $y$.  It should be noted that, by virtue of the discrete 
nature of the Markov chain, time in this context is measured by 
the number of mutation cycles experienced by the system.  The 
probability we are looking for takes the form
$$\peemu \left(\inf_{i \leq n} \tau_i (M) \leq y \right) = 1- 
\left( 1- \peemu \left(\tau_1 (M) \leq y \right) \right)^n,$$
where $\peemu$  denotes the path measure induced by the Markov 
chain and the index $i$ tallies the replica under consideration.  
Since we are focusing our attention to the GC reaction, $n$ is 
approximately $10^3$-$10^5$.

\section{Methodology}

The study of the evolutionary optimization process outlined in the 
previous section uses results by the second author on the 
covergence rates of exit times of Markov chains \cite{ted1}.  
The general approach for estimating the desired 
tails of the exit time distributions consists of the following 
steps:

\begin{itemize}
\item [{(i)}] 
We formulate a Dirichlet problem for $\gee$ on $\finv \minf$ whose 
solution provides a martingale representation of the Laplace 
transform of the exit time $\tau(M)$.

\item [{(ii)}] 
We solve the resulting Dirichlet problem and compute the desired 
Laplace transform as
$$\psi(\xi) \stackrel{\Delta}{=} \avmomgen = {\frac {\left(1- 
pe^\xi \right) \sumj^b q(j) \pjejx}{1- e^\xi +
(1-p) e^\xi \sumj^b q(j) \pjejx}}$$
where $\ee^\ast$ denotes the expectation starting from a 
$\mu$--distributed initial sequence.

\item [{(iii)}] 
We compute the Legendre-Fenchel transform $I(y)$ of the cumulant 
of $\tau(M)$ as
$$I(y) = \left\{ \begin{array}{ll}
\int_{1}^{\frac {y}{\ee^\ast [\tau]}} \Xi(t) dt \cspace & \mbox{if 
$y \geq \ee^\ast [\tau]$}
\\ 
\int^{1}_{\frac {y}{\ee^\ast [\tau]}} \Xi(t) dt \cspace & \other
\end{array} \right. ,$$
where $\Xi(t)$ is the (positive or negative depending on whether 
$y \geq \ee^\ast [\tau]$ or not) solution to 
$${\frac {d \psi}{d \xi}} \left( {\frac {\Xi(t)}{ \ee^\ast 
[\tau]}} \right) =t\ee^\ast [\tau] \psi \left( {\frac 
{\Xi (t)}{ \ee^\ast [\tau]}} \right).$$
It turns out \cite{ted1} that, for $y \leq \ee^\ast 
[\tau]$,
$$\peemu \left(\tau_I (M) \leq y \right) \approx \exp \left\{ I(y) 
\right\}.$$

\item [{(iv)}] 
We estimate $\Xi(t)$ by performing a Taylor expansion of the 
cumulant of $\tau(M)$ at $-\infty$, yielding
$${\frac {d \psi}{d \xi}} \left(\log \lambda \right) 
=\sum_{i=1}^\infty {\frac {c_i \lambda^i}{i!}},$$
as $\lambda \searrow 0^+$.  Inverting the polynomial on the right-hand side we obtain the general form $$\Xi(t) \approx c_1 \log 
\left(c_2 t \right).$$
Specifically, if we stop the Taylor series after the linear term 
we have $c_1=1$ and $c_2={\frac {p+\beta (1-p) \left[1- q(0) 
\right]}{\beta q(0)}}$.
\end{itemize}

\section{Results}

Our goal in the modeling exercise described in the previous few 
sections was not to produce exact quantitative predictions for the 
behavior of the immune system.  Instead, our main goal has been to 
elucidate the drivers behind the apparently complex behavior of 
the immune system and develop a qualitative understanding of what 
makes the system work so efficiently.  With this in mind, we 
acknowledge that our model is purposely kept simple enough to 
allow a thorough analysis and simulation.  

Our results are presented in this section in order of increasing 
complexity.  First we discuss the surprisingly quick response of 
the system which is governed by a variant of the recently 
discovered {\it cutoff phenomenon} in many Markov chains.  Then, 
we proceed to exhibit the dependence of the system's response time 
on the value of the trade-off parameter $p$.  A relatively narrow 
band of values for $p$ are shown to significantly outperform all 
others.

Until this point, we consider one antigen and resulting 
affinity landscape at a time.  In order to appreciate the general 
applicability of the trade-off discussed above, we proceed next to 
describe the response dynamics as a function of $p$ for a sequence 
of landscapes, randomly generated from the biologically motivated 
range of parameters presented earlier.  This investigation yields 
a sharp transition between two regimes: a frozen, ordered regime 
with too little randomization and a liquid, chaotic regime with 
too much randomization.  The corresponding {\it phase transition} 
occurs within the same range of $p$ that strikes the right balance 
in the trade-off studied before. 

The next step is to investigate the trade-off between the 
diminishing incremental benefit in affinity against the increasing 
time  burden as the system attempts to reach higher affinity 
peaks.  This trade-off may cast some light on the mechanism that 
ends the hypermutation process.

Once we have established the existence of the narrow band of 
desired values of the parameter $p$, we finally turn our attention 
to its biological implementation.  How does the system know to set 
the parameter $p$ at the right level?  It turns out that it 
doesn't need to know.  In fact, with some mild assumptions, we 
show that the system cannot help but adapt to exhibit a value of 
$p$ within the narrow desired range despite exogenous shocks.

\subsection{The Cutoff Phenomenon}

\begin{figure}
\epsfxsize=4in
\epsfbox{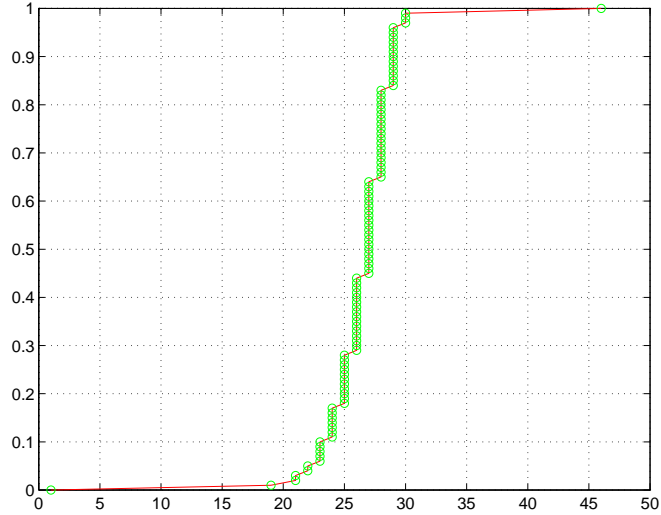}
\caption{Convergence of $\peemu \left(\tau(M) \leq y \right)$ as a 
function of $y$}
\label{fig:cutoff}
\end{figure}

This phenomenon has been studied by Diaconis and his collaborators 
in \cite{diaconis1,diaconis2,diaconis3}.  In the 
context of the Markov chain modeling the affinity maturation 
dynamics, Fig. \ref{fig:cutoff} shows a typical realization of 
the observed cutoff.  Methodologically, our approach differs from 
that of Diaconis in the measure of convergence used.  The more 
traditional approach employs the total variation distance between 
the sample distribution of the Markov chain after a finite number 
of steps and the stationary distribution.  Instead, we concentrate 
our attention on the left tail of a family of exit times.  At a deeper level, 
the two approaches are not as dissimilar as they may 
appear.  The technique for estimating the total variation distance 
often relies on coupling arguments which reduce to the 
distribution of the coupling time, a stopping time not unlike the 
ones underpinning our approach.

In the immunological literature, this behavior has been captured by
previous models of the humoral response maturation process
\cite{celada,kepler,oprea}.
Qualitatively, this result corroborates the observed speed of the 
affinity maturation in the immune system.  Within a few mutation 
cycles spanning 3--4 days, the specific immune response gains 2--3 
orders of magnitude in affinity.  This time, measured in the number 
of mutational cycles, is at least 2 orders of magnitude less than 
the expected time to equilibrium of the Markov chain.  Thus, the 
observed affinity maturation is a decidedly disequilibrium effect.  
This is certainly one of the sources of complexity in the system 
that defies interpretation using traditional equilibrium-oriented 
techniques.

\subsection{The Optimal Value of $p$}

\begin{figure}
\epsfxsize=4in
\epsfbox{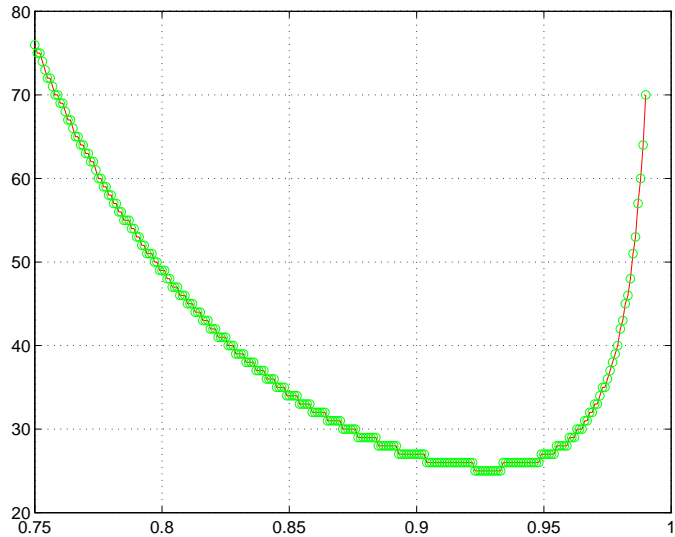}
\caption{$\tau(M)$ as a function of $p$}
\label{fig:trade-off}
\end{figure}

The result described in this section is an outgrowth of previous 
work by the second author.  In \cite{ted3} similar techniques as 
the ones described above were applied in the study of the 
asymptotic convergence rate of a class of Markov chains 
encompassing the one employed here.  In that context, it was shown 
that under very mild conditions on the landscape, there is a 
nonzero level of {\it randomization by design} (i.e., beyond the 
minimum randomization required to avoid remaining trapped in 
strictly local optima asymptotically) that substantially increases 
the convergence rate.  

Here we study the more complex problem of transient behavior of 
the Markov chain.  Nevertheless, we obtain a qualitatively 
equivalent result.  A fine-tuning of the parameter $p$ in a narrow 
range of values confers a remarkable performance improvement 
(Fig. \ref{fig:trade-off}).  This qualitative behavior appears 
to be surprisingly insensitive to the structure of the landscape.  
In \cite{ted3} this was investigated for exponential landscapes 
similar to the ones studied here, as well as broader classes of 
polynomial, logarithmic, and uniformly random landscapes.  In all 
cases, the optimal level of $p$ was within a narrow overlapping 
range.  It is this hint of universality that has led us to the 
robustness investigation that follows next and which uncovers a 
further charactreristic of this class of complex systems.

\subsection{Performance Robustness}

\begin{figure}
\epsfxsize=4in
\epsfbox{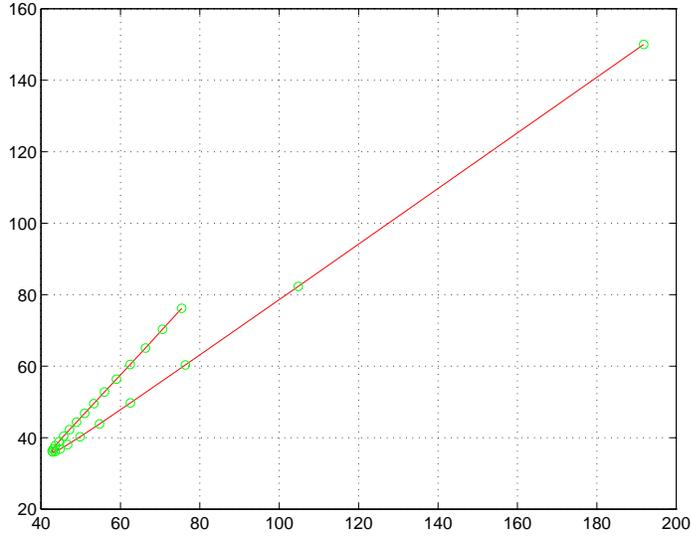}
\caption{Expected value of $\tau(M)$ versus standard deviation of 
$\tau(M)$}
\label{fig:robust}
\end{figure}

As alluded to above, performance robustness refers to the 
surprising consistency in the efficiency of the immune system 
response to a combinatorially large set of invading antigens.  We 
studied this problem by performing the trade-off analysis for the 
optimal value of $p$ for a population of randomly generated 
landscapes from within the biologically justifiable range of 
parameters mentioned earlier.  One thousand different antigens 
were tried, and the resulting mean and standard deviation of the 
response time were plotted for a series of $p$ values.  In Fig. 
\ref{fig:robust} we have suppressed the third dimension ($p$) in 
order to illustrate more clearly the two observed regimes.  For 
clarity, note that $p=0.75$ corresponds to the point close to 
(80,80) in the graph, while $p=0.99$ corresponds to the point 
close to (190,150), with equally-spaced, increasing $p$ values in 
between.
We observe that for very high values of $p$, the system's expected 
performance suffers a rapid decrease accompanied by an increase in 
the variability of that response across different antigens.  As we 
lower the $p$ value, there is a narrow range between about 0.85 
and 0.91 for which the system attains the best response time and 
the lowest variability of that performance as it faces varying 
antigens.  Past that point, there is a sharp change in the 
system's behavior.  The expected response time deteriorates and 
the variability of that response grows even faster.  
Using the same approach as in \cite{ted3}, we identify the first 
of these two regimes with a {\it solid} phase which is too ordered 
to escape the strictly local optima that abound in a randomly 
generated ladnscape.  Similarly, the second regime is seen as a 
{\it liquid} phase with too little structure to effect the desired 
progression towards higher affinity peaks.  Instead, systems in 
this regime appear to diffuse aimlessly in sequence space.  The 
observed sharp transition between these two regimes is analogous 
to the concept of {\it the edge of chaos} introduced by Kauffman 
\cite{kauffman} as well as the notion of a critical level of 
parallelism investigated in \cite{siapas}.

\subsection{The Affinity Threshold}
\label{bioaffinity}

\begin{figure}
\epsfxsize=4in
\epsfbox{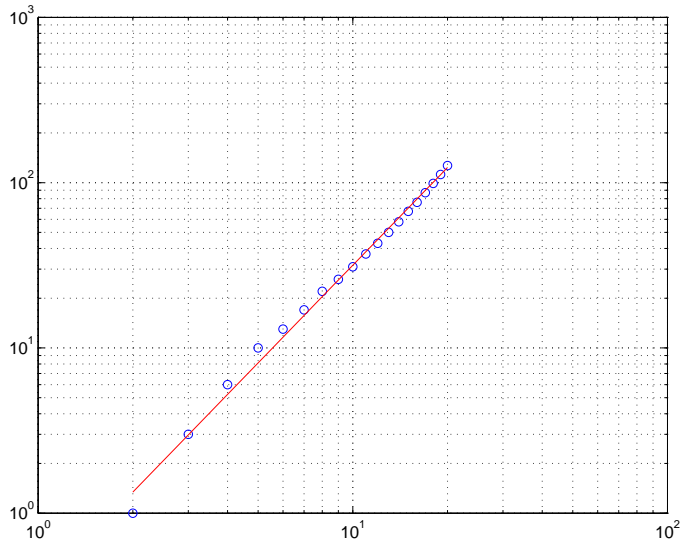}
\caption{$\tau(M)$ as a function of the desired affinity threshold 
$M$}
\label{fig:threshol}
\end{figure}

So far, our results related to the time it takes the system to 
achieve a fixed desired level of affinity.  In this subsection we 
ask the question how the system knows when to terminate the 
hypermutation process.  Clearly, a new trade-off becomes relevant 
between the energetic costs and immunologic risks (e.g. 
autoimmunity or malignancy) of continuing the hypermutation 
process for longer and the expected incremental affinity gains as 
discussed in detail in Sect. \ref{bioaffinity}.  

\begin{figure}
\epsfxsize=4in
\epsfbox{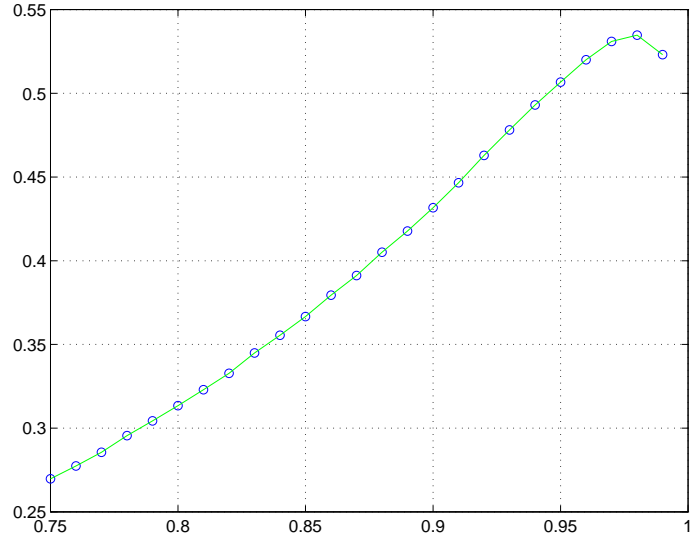}
\caption{$\alpha$ as a function of $p$}
\label{fig:alpha}
\end{figure}

As one might expect, diminishing incremental affinity gains are a 
rule.  Figure \ref{fig:threshol} exhibits the power-law 
relationship that holds between the response time and the 
resulting affinity for a fixed value of $p$.  We are interested to 
investigate how this power law varies as $p$ moves across the two 
regimes outlined above.  In general, we obtain an approximate 
power law of the form $M(\tau) \sim \tau^\alpha$.  Figure 
\ref{fig:alpha} shows the behavior of $\alpha(p)$.  We observe 
that the rate at which incremental gains diminish increases as $p$ 
is lowered deeper into the liquid phase.  It is worthwhile 
to note that the value of $\alpha=0.5$, which corresponds to the 
scaling behavior of Brownian motion, is attained close to the 
value of $p$ that leads to the phase transition described in 
the previous paragraph.

One biological interpretation of this observed variation of the 
scaling law as a funtion of $p$ is that lower values of $p$ are 
intrinsically riskier, in that they entail more global 
jumps.  Thus, it is not surprising that lower values of $p$ 
exacerbate the trade-off between time and diminishing affinity 
gains in favor of stopping the hypermutation process sooner.  We 
conjecture that this type of behavior leads to a two-tiered 
response:  

\begin{itemize}
\item [{(i)}] 
After a relatively modest amount of time, a sufficient affinity 
improvement has been gained to allow the mature plasma cells to 
exit the GC and mount a highly specific and effective attack on 
the invading antigen.

\item [{(ii)}] 
The hypermutation process continues in the background and now 
strives to generate an even more specialized population of memory 
cells to maintain long-term immunity to the antigen in subsequent 
reinfections.
\end{itemize}

The above results lead to the consideration of the repertoire shift
phenomenon.  The emergence of this phenomenon has been studied in
earlier models.  Specifically, within the population dynamics
paradigm, Shannon and Mehr in \cite{ramit} use the ``destructive
effect of hypermutation on the primary B cell repertoire'' to show
that the memory cell population converges to a moderately high
affinity level: lower than the constituents of the fast primary
response, which die out due to the overwhelming likelihood of
affinity-reducing (and thus lethal) subsequent mutations, but higher
than the background average due to the hypermutation and
antigen-constrained selection processes in the germinal centers.

The global optimization paradigm presented in this paper offers a
further, direct explanation for the occurrence of repertoire shift.
While the result is not new, we present it mainly to illustrate the
usefulness of the proposed paradigm.  During the ongoing hypermutation
process suggested in step (ii) above, new peaks in the affinity
landscape eventually emerge, arising from V genes not present in the
primary response.  Specifically, while some sufficiently high affinity
peaks are reached relatively fast during the primary response, they
sometimes lead to lineages whose affinity plateaus shortly thereafter.
At the same time, other clones, which found themselves in more
modestly rising affinity ``hills'' during the primary response,
eventually experience more rapid affinity improvement, which leads them
to surpass the performance of the lineages that dominated the primary
response.

\subsection{Evolving to the Edge of Chaos}

Having exhibited the critical dependence of the response time to 
the value of $p$, we address the question of how the immune system 
knows to fine-tune the value of $p$ within the narrow desired 
band.  We hypothesize that the nature of the evolutionary dynamics 
is such that the system cannot avoid being {\it drawn} to operate 
within the desired band of $p$ values.

\begin{figure}
\epsfxsize=4in
\epsfbox{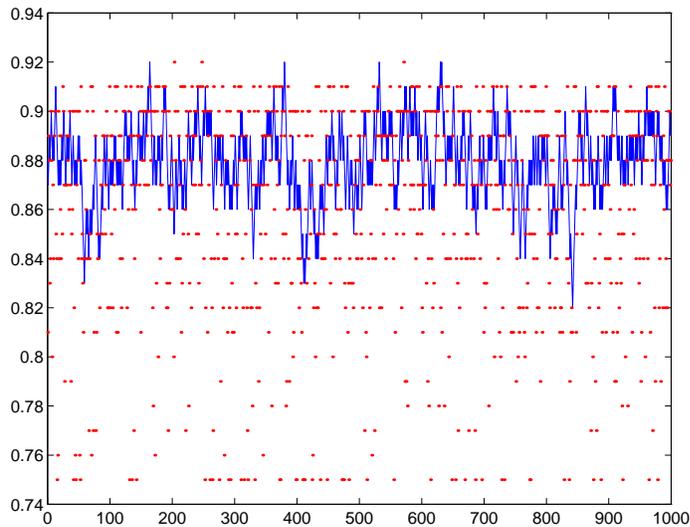}
\caption{Stability of evolving $p$ inside robust band}
\label{fig:evolve1}
\end{figure}

In order to examine this hypothesis, we conjectured that the 
mutator control mechanism which biologically instantiates the 
parameter $p$ is itself coded in the genome along with the rest of 
the Ig molecule.  Even though that part of the genome is not 
involved directly in the affinity-based selection as it doesn't 
code for the Ig molecule, it may be subject to a longer scale mutation and selection process.  
Such a stipulated evolution may theoretically occur at the level of individual
cells (as is the case with the evolution of Ig affinity during the hypermutation process),
or at the level of organisms across more traditional evolutionary time scales. The
mechanics of mutation control, which account for the $p$ value, are not understood well enough to determine this point with certainty.  In any case, the simulations we present below of an evolving $p$ value apply equally well to either interpretation. The idea is that, as a population of individuals is facing ever changing antigens, $p$ values in the optimal range dominate the population. The main difference between the two interpretations would be in the meaning of individual (cell vs. organism).
For the rest of this subsection we assume that such a process exists.  We further 
assume that we can model the process as a reversible nearest 
neighbor random walk with biased transitions towards the 
neighboring $p$ value that reduces the immune response time.  We 
justify this bias by the strong selectional advantage of clones 
with faster immune response times.
Using this hierarchical evolution model, we simulated one thousand 
successive infections with different antigens.  During each 
infection the system has a fixed value of $p$ and behaves as 
described in the sections above.  Between infections, the system 
performs one step of the biased random walk, motivated by the 
selectional advantage of the clones with the faster response time 
to the previous infection.
Fig. \ref{fig:evolve1} depicts the evolving value of $p$ in the 
population (solid line) and the optimal value of $p$ corresponding 
to the landscape for each new infection (dashes).  This optimal 
$p$ value for each new landscape acts as an exogenous shock to the 
system.  Figure \ref{fig:evolve1} shows that despite these 
periodic exogenous shocks (which would have the system decrease 
its $p$ value into the liquid phase), the system stays 
within the desired narrow band, without explicit instruction.  

\begin{figure}
\epsfxsize=4in
\epsfbox{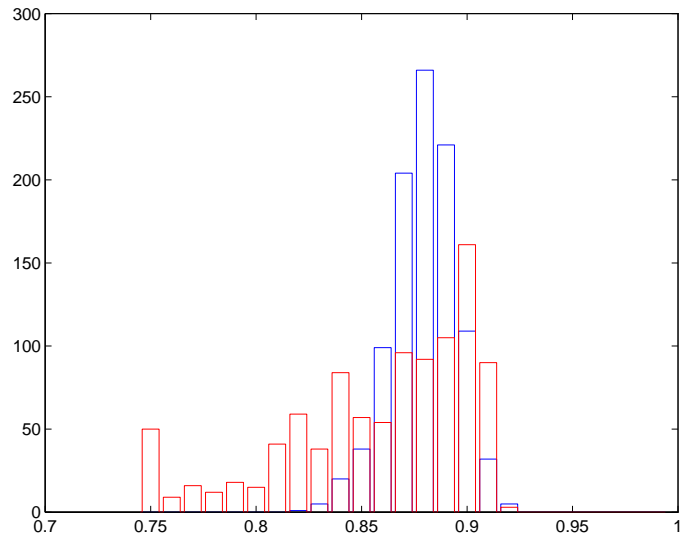}
\caption{Comparison of the distribution of evolving $p$ versus the 
distribution of optimal $p$ for the sequence of landscapes that 
drive the system}
\label{fig:evolve2}
\end{figure}

\begin{figure}
\epsfxsize=4in
\epsfbox{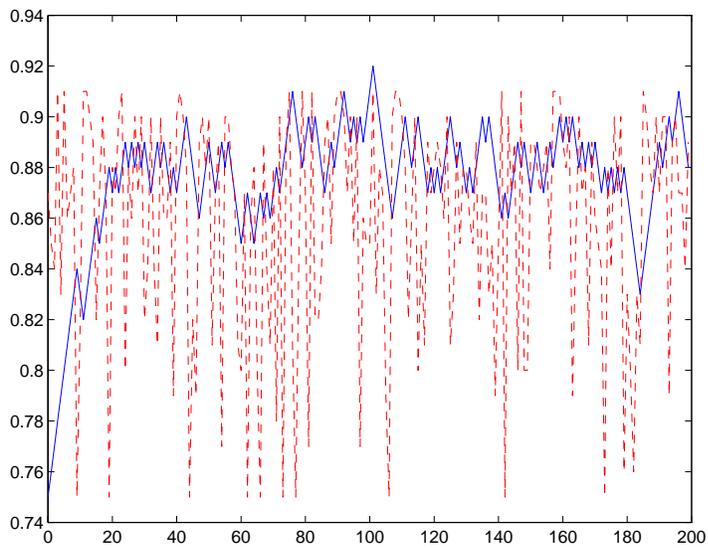}
\caption{Convergence of evolving $p$ to robust band}
\label{fig:evolve3}
\end{figure}
In Fig. \ref{fig:evolve2} we show the histograms associated with 
the $p$ values in Figure \ref{fig:evolve1}.  Once again, we see 
the system converging to a much tighter distribution of $p$ values 
than the exogenous shocks.  Finally, Fig. \ref{fig:evolve3} 
shows the transient behavior of this hierarchical evolution model.  
We purposely started the system at $p=0.75$, distinctly outside 
the desired range, and let it evolve autonomously.  Once again, 
without any instruction, the system converges rapidly to the 
desired band of $p$ values and is stabilized within that band.

\section{Conclusions and Directions of Further Study}

Our model gives evidence that the mutator mechanism functioning 
during somatic hypermutation has evolved to a trade-off value of 
$p$ that gives a fast, efficient, and consistent response.  While this
mechanism is still not understood, certain transcriptional 
elements appear to be necessary for mutation in  both heavy and 
light chain genes, although additional, novel molecular mechanisms 
need to be considered.  Our model shows that {\it global jumps} 
need to occur during the process, and from a biologic standpoint, 
these may be understood in two ways.  They may be encompassed by 
the less frequent, but necessary mutations in the more diverse 
regions of the CDRs or in the FRW regions, or in the form of 
deletions and insertions.  This latter characteristic in 
particular implies the occurrence of double strand breaks in the 
DNA and may support evidence of recombination \cite{kong} or 
reverse transcription models \cite{blanden}.  

With regards to the affinity threshold, our model demonstrates the 
diminishing incremental gains in affinity as a function of an 
increasing number of mutations.  Future investigation might 
inlcude analysis of the selection process during the ongoing 
shaping of the memory compartment that likely encompasses 
alternative selection parameters, kinetics, for example, being one 
of them \cite{foote2}.  

Our model also shows that the evolution of mutations seems to have 
an internal driver that points us toward the higher order template 
underlying the observable hotspots.  This finding emphasizes the 
evolutionary efficiency with which nature approaches this problem.

As was mentioned during the description of the model, the process 
of hypermutation has been suggested to be redundant or unnecessary 
\cite{blanden}.  We proposed larger control mechanisms 
underlying the changes between primary and secondary repertoire 
that might explain the persistence of this process.  A next step 
might include an attempt to understand these mechanisms, perhaps 
in regards to tolerance and early development, or in regards to 
the dynamics of the antibody repertoire acting beyond affinity 
selection.  Hypermutation may be a necessary step to set the stage 
for these other processes to occur.

\bibliographystyle{plain}
\bibliography{immune}

\end{document}